\documentclass[12pt]{article}
\usepackage{amsmath,amsthm, amssymb}
\usepackage{graphicx}
\usepackage{leqno}
\usepackage{hyperref}
\usepackage{setspace}
\numberwithin{equation}{section}
\begin{document}
\title{\vskip-40pt       Relativity Is Not About Spacetime }
\author{Edward J. Gillis\footnote{email: gillise@provide.net}}
				
\maketitle

\begin{abstract} 

\noindent 

  Quantum measurement predictions are consistent with relativity for macroscopic
observations, but there is no consensus on how to explain this consistency in
fundamental terms. The prevailing assumption is that the relativistic structure
of spacetime should provide the framework for any microphysical account. This
bias is due, in large part, to our intuitions about local causality, the idea 
that all physical processes propagate through space in a continuous manner. 
I argue that relativity is not a guarantor of local causality, and is not
about ontological features of spacetime. It is, rather, an expression of the 
observational equivalence of spacetime descriptions of physical processes. 
This observational equivalence is due to the essentially probabilistic nature of 
quantum theory.

\end{abstract}

\section{Relativity Is Not Based On Local Causality}

	In order to get from here to there, you have to pass through points in 
between. This intuitively obvious notion is fundamental to the  
implicit model of the external world that shapes our thoughts and actions. 
Other vertebrate species also exhibit behavior that appears to be guided 
by analogous cognitive processes. I recall watching the children of some 
friends  use a laser pointer to tease their cat. They would shine the laser 
spot on the floor in front of the cat, move it around enough to get the 
cat's attention, and induce her to start tracking it. They would then 
jerk the spot of light to a completely different place in the room, 
leaving the pet totally bewildered after witnessing a seemingly coherent 
object jump discontinuously through space.

The expectation that things move through space in a continuous fashion is 
very deeply ingrained. Developmental psychologists have attempted to determine 
to what extent it might arise from innate neural processes.\footnote{See, 
for example, \textit{Spatiotemporal continuity and the perception of causality in infants}, 
by Leslie\cite{Leslie}.}   
No matter what one thinks about the relative contributions of heredity and 
learning in generating this belief, it clearly plays an enormous role in 
guiding our everyday life and in the ways that we think about physics.

Its place in our world view and its influence on the rise of modern science 
are evident in Newton's efforts to justify and reconcile gravitational 
action at a distance with the received belief that causes are transmitted 
by contact through some medium. It is not completely certain how he thought 
that the reconciliation was to be achieved, but he clearly acknowledges the 
prevailing notion in his 1693 letter to Bentley\cite{Newton_letter}:
\begin{quotation}\noindent "...That gravity should be innate, inherent, and essential 
to matter, so that one body may act upon another at a distance, through a vacuum, 
without the mediation of anything else, by and through which their action and 
force may be conveyed from one to another, is to me so great an absurdity, that 
I believe no man who has in philosophical matters a competent faculty of thinking, 
can ever fall into it."\end{quotation}

The success of Newton's theories of mechanics and gravitation led people to accept 
the idea of action at a distance, at least provisionally. However, the intuition 
that all processes must propagate in a continuous manner is so compelling that the 
principle of local causality gradually reasserted itself in physics. The concept of 
gravitational potential, employed by Euler, Lagrange, and Laplace hints at the idea 
of a field defined at all points in space. In the nineteenth century, the work of 
Faraday and Maxwell fully established the idea of continuously propagating electromagnetic 
fields, and suggested to some the existence of a luminiferous ether. It also led to 
a clear conflict with Newtonian mechanics, which was resolved by Einstein's Special 
Theory of Relativity.

Einstein's theory revolutionized our ideas about time, with radical consequences for 
the concepts of mass, motion, and energy. However, it also reestablished physics on 
the venerable principle of local causality. In this respect, it was 
a very conservative revolution. 

Einstein apparently cemented the triumph of this traditional principle with 
the development of the General Theory of Relativity, which is able to explain the 
propagation of gravitational influences in a locally causal manner.

Quantum mechanics was developed at roughly the same time as the theories of relativity. 
Although he made major contributions to its development, Einstein was unhappy 
with some of the central features of quantum theory, in particular, its 
probabilistic nature.  In his paper with Podolsky and Rosen\cite{EPR} he argued 
that some of the predicted correlations for entangled systems would 
require 'spooky action at a distance', if one assumed that the effects were truly 
nondeterministic. The conclusion drawn in the EPR paper was that quantum mechanics 
needed to be 'completed' by a fully deterministic theory. 

Bell turned the EPR argument around\cite{Bell_1,Bell_2}. He showed that the correlations 
between some specific pairs of outcomes of spacelike-separated measurements on 
entangled particles are too large to be explained by the assumption that the results 
are determined by some factors in the common past of the measured systems.\footnote{Aside 
from superdeteministic explanations.} These nonclassical "super" correlations appear to 
indicate that the particular result of one of the measurements acts across a spacelike 
interval to change the probabilities of the outcomes of the other measurement. There are 
real superluminal effects.
		
		So why cannot these effects be used to send superluminal signals? The reason is that 
one cannot control the outcomes of the local measurement. They are \textit{essentially 
probabilistic}. When all of the possible results are summed over, the total probability 
of any particular distant outcome is unchanged. It is a striking result that the Born 
probability rule\cite{Born}, and \textit{only the Born probability rule}, is able to 
maintain this consistency. Any slight deviation from it would enable superluminal information 
transmission. It is not hard to see this, given a sufficient number of identically prepared 
entangled systems. This is, of course, also a consequence of Gleason's theorem\cite{Gleason}.

This situation is sometimes described as demonstrating the "peaceful coexistence" of quantum 
theory and relativity\cite{Shimony_1,Shimony_2}. In a number of commentaries, it even seems 
to take on a rather miraculous air. But, is it not much more straightforward to assume that 
the impossibility of sending superluminal signals, and, \textit{hence, relativity}, is a 
consequence of the fundamentally probabilistic nature of the elementary interactions that 
constitute measurements?

Note that the same line of argument that shows how the Born rule prevents superluminal 
signalling also shows that it is impossible to say which of the two (or more) 
measurements is affecting the other. In other words, there is no \textit{observable} 
sequence of these spacelike-separated events. This is one of the fundamental characteristics 
of the relativistic description of spacetime. 
  
Einstein and his colleagues set out to show that if physical theory retained its 
nondeterministic elements, it would have to countenance nonlocal actions. Bell 
showed that the inclusion of nonlocal effects is unavoidable, and subsequent 
analysis\cite{Elitzur_92,Popescu,Elitzur_2,Masanes} has demonstrated that the 
nondeterministic nature of these effects is essential to the prevention of superluminal 
signalling. It is ironic that relativity is preserved by the very property of quantum 
theory that Einstein found so disturbing - indeterminism.

\section{Making Sense of Signal Causality}

We have misunderstood what relativity is really about. This is a result of
the historical path to its discovery, our pretheoretic biases, and the
elegance of the theory. Its two great classical instantiations, Maxwell's
electromagnetism and general relativity, capture our intuitions about
continuous propagation so beautifully  that we perceive local causality as
a manifestation of the metric structure of spacetime. This perception
fosters an attitude that, no matter how serious a problem is, any proposed
solution should fit comfortably within this structure.\footnote{At least 
until we approach the Planck scale} However, even without taking into account 
the challenges posed by quantum theory, there are reasons to question the
completeness of the classical relativistic world view. The infinite self-energy
and radiation reaction problems of electrodynamics have provoked proposals
that violate conventional notions of causality\cite{Feynman}. Solutions of 
Einstein's gravitational equation that approximately describe our universe 
exhibit generic singularities, where spacetime and the laws of physics break 
down\cite{Hawking-Penrose}. Since physicists have worked on these issues for 
roughly a century, one might start to wonder whether there are inherent logical 
limits to the notion of local causality, analogous to the limits on provability in
mathematics.\footnote{This is only an analogy. I am not suggesting that quantum 
indeterminism is logically linked to Goedel's incompleteness theorem} 

In any case Bell's analysis has forced us to recognize that not everything 
that happens in our world can be explained in terms of locally propagating 
processes. The fact that local causality is not the basis for contemporary 
physics is often obscured by a failure to distinguish between it and the 
weaker, more general notion of signal causality, the prohibition of superluminal 
information transmission. Local causality implies signal causality, but signal 
causality permits superluminal effects, provided that they are nondeterministic, 
and in accord with the Born rule. 

This more general principle is implemented in quantum field theory as the requirement 
that spacelike-separated field operators commute\cite{Norsen}. This means that a measurement 
made in one location cannot affect the overall probabilities of possible outcomes 
of a measurement made in a spacelike-separated region, even though specific 
pairs of outcomes exhibit correlations that indicate some kind of linkage across 
the spacelike interval between them. 

Most quantum field theory texts describe the commutativity requirement as a causality 
condition, usually without qualifying it as \textit{signal} causality. In his 
book, Weinberg\cite{Weinberg} adopts a somewhat different perspective: 
              \begin{quotation} 
     		{"The point of view taken here is that [the commutativity requirement] is needed 
     		for the Lorentz invariance of the S-matrix, without any ancillary assumptions 
     		about ... causality."} \end{quotation}		
One might say that while most authors take the requirement as a formal expression of 
causality, Weinberg prefers to view it as preserving relativity. These complementary 
interpretations highlight the intimate connection between the two principles. It must be 
emphasized, however, that the attribution of Lorentz invariance to the S-matrix does 
\textit{not} automatically assign some special ontological status to the Lorentz-invariant 
metric of Minkowski spacetime. The S-matrix is a set of theory-laden calculations, and 
it reflects a number of properties of the mathematical formalism. If the metric properties 
of spacetime alone were sufficient to guarantee the Lorentz invariance of observable 
quantities, there would be no need for a special postulate to govern the action of 
field operators. This additional assumption is aimed at regulating the nonlocal effects 
that are implicit in quantum theory.
          
The S-matrix calculations yield probabilities for measurement outcomes. These quantities 
are compared to readings of macroscopic laboratory instruments, and the meaning of the 
commutativity requirement in relation to them is clear. However, problems arise 
when we try to translate the commutativity condition into a statement about 
\textit{fundamental physical processes}. 

This translation problem derives from the way in which contemporary physical theory 
is structured. Dynamic equations describing elementary processes are completely 
deterministic. Nonlocal, probabilistic effects have been banished to a lawless, 
ill-defined border region that lies somewhere between microscopic and macroscopic realms. 
This theoretical structure has been incredibly fruitful, but it has also made it 
incredibly difficult to understand what it is saying about the world at the most 
fundamental level.

This is why Bohr thought that references to macroscopic, classical concepts are essential 
to interpret statements about microphysical processes\cite{Bohr_1,Bohr_2}. Bell described 
the problem very eloquently\cite{Bell_LNC}: 
\begin{quotation}\noindent "More importantly, the 'no signalling...' notion rests on concepts
which are desperately vague, or vaguely applicable. The assertion that 'we cannot signal faster 
than light' immediately provokes the question:

 Who do we think \textit{we} are? 
 
 \textit{We} who can make 'measurements'...?... chemists, or only physicists,...pocket 
 calculators, or only main frame computers? "\end{quotation}   
Bell raised the possibility that the relation of signal causality to elementary processes  
is analogous to that of thermodynamics, but he did not think that this was a promising 
approach.

Maudlin\cite{Maudlin} has extended Bell's critique. After acknowledging that the notion 
of 'signalling' apparently makes essential reference to human activity, he demonstrates 
that the Born probability rule precludes the use of nonlocal quantum effects for this 
purpose. However, he goes on to argue that the nonlocal changes in the states of elementary 
systems do imply some sort of superluminal transmission of information to \textit{that 
particular} system, even though the information is not, in general, accessible to any 
other physical system.

The criticisms of Bell and Maudlin have considerable merit, and they pose a serious problem 
for the concept of signal causality. It appears that when we formulate it in terms of 
elementary physical processes (rather than human activities) as a prohibition of 
superluminal information transmission, it does not really apply to the kinds of nonlocal 
quantum effects that we are trying to describe. 

To deal with this problem I will try to show that, despite Bell's reservations, the 
relationship of signal causality to elementary processes \textit{is} analogous to that 
of thermodynamics by noting some important parallels between the concepts of 'temperature' 
and 'information'. The concept of information that I will use is distinct from that of 
Maudlin. It will allow us to define signal causality as the prohibition of superluminal  information transmission, and it will depend in an essential way on the nondeterministic 
nature of elementary processes.

When we say that the temperature in a room is $20^o$ C, we are implying a great 
deal about conditions in the room. For example, the room is in thermal equilibrium - 
no significant net transfers of energy will take place between adjacent masses of 
air. Individual molecules do not have temperatures. They can have almost any energy, 
and in a collision with another molecule there is an overwhelming likelihood that 
a substantial amount of energy will be exchanged. The concept of temperature 
applies to very large collections of molecules that meet the proper conditions.
Of course, the term 'temperature' is sometimes applied to individual elementary 
particles as a synonym for 'energy'. There is an important link between 
energy and temperature, but, as just noted, the ascription of a definite temperature 
implies a number of relationships among physical systems.

I maintain that Maudlin's application of the term 'information' to individual  
elementary systems is analogous to ascribing a temperature to a single molecule. 
What Maudlin means by 'information' is a full specification of the parameters 
required to characterize an elementary state. This is a perfectly reasonable 
use of the term, depending on the circumstances, just as talking about the 'temperature' 
of individual particles makes sense in certain contexts. But, as Maudlin emphasizes, 
this 'information' is not accessible to any other physical systems. 

The concept of information is, to a large extent, relational. Physical systems that 
can instantiate information must be able to represent the states of other systems, 
and to transmit and receive the relevant types of representations. Maudlin's inaccessible 
information clearly lacks these attributes.

To obtain information about the state of a physical system one induces it to interact 
with other particles in such a way that correlations are established between the states 
of the subject and detector systems. A chain or cascade of correlating interactions is 
set up so that the state of the original subject can be represented in a much larger 
system. If these interactions were completely deterministic and reversible, one could 
generate a complete description of the original state. The reason that we cannot, in 
general, obtain such a description is that the elementary interactions that establish 
correlations between subject and detector particles \textit{are not completely deterministic}.

If we explicitly recognize the nondeterministic behavior at the level of elementary 
processes,  we can understand why information cannot be transmitted superluminally. 
Individual elementary particles can possess a definite state, but they cannot, by 
themselves, instantiate information about that state because interactions with other 
particles can change the state in a nondeterministic way. There would then be no faithful, 
physical representation of the original state. This is the content of the No-cloning 
theorem\cite{No_Clone}. The preparation of individual particles in definite, known 
states requires interaction with very large numbers of other particles. It is only at 
this scale, through the full network of correlations that are established, that  
information can be said to exist. Information about a physical system must be fully 
accessible to other systems, and it must be reasonably stable against probabilistic 
changes. This means that the definition of physical information is a matter of gradation 
or degree; but it is still a perfectly objective concept. It is analogous to temperature, 
which is perfectly well-defined, even though we do not precisely specify how big a system 
must be to achieve thermal equilibrium.

The point is that the limitations on information and on its transmission are 
explainable in terms of a basic property of elementary interactions - their 
probabilistic nature. As stated above, this issue has been clouded by the fact 
that contemporary theory does not acknowledge any deviation from complete determinism 
at this level. There are a number of reasons for this, but the biggest one is 
that such an acknowledgement completely disrupts our understanding of spacetime 
structure.

\section{The Status of Spacetime Structure}

It is possible to describe nondeterministic behavior at the level of elementary 
interactions, at least in a phenomenological way. The linearity of the Schr\"{o}dinger 
equation permits nonlocal, probabilistic transfers of amplitude between interacting 
and noninteracting branches of the wave function without violating signal causality. 
The details have been given elsewhere\cite{Gillis_E}. The catch is that these 
hypothesized effects are both nonlocal and nondeterministic. Nondeterminism entails 
irreversibility, and irreversibility implies that the effects must be sequenced. If 
interactions involving the same entangled system occur in spacelike-separated regions, 
there is no way to account for the sequencing of the nonlocal effects associated with 
the interactions simply by reference to the relativistic metric structure. 

	This apparent conflict with relativity is present in conventional descriptions 
of quantum measurement. To see this assume that the wave function of a single particle 
has bifurcated into two main branches, and that there is an ideal detector in the path 
of each branch. One of the detectors registers the presence of the particle; the other 
does not. Suppose that the two measurements are spacelike-separated. In some reference 
frames the positive outcome occurs first; in others the negative outcome occurs first. 
The two different sequences are associated with two distinct accounts: (1) the positive 
outcome collapsed the wave function, setting the amplitude of the alternate
branch to zero, insuring a negative outcome for the other measurement; (2) the
negative outcome collapsed the wave function so that all of the amplitude was
concentrated in the other branch, insuring a detection.

Each of these sequences consists of a nondeterministic, irreversible event followed 
by a deterministic, reversible one. The problem is that the order is switched, with 
the positive outcome first in one account, and the negative outcome first in the 
other\footnote{Treating the events as simultaneous does not work. See \cite{Gillis_E}.}. 
The descriptions are \textit{observationally equivalent, but logically incompatible}. 
To deal with this conundrum many try to interpret quantum wave functions in a purely 
epistemic way, but this approach is fraught with difficulties. 

Conventional theory deals with this conflict by completely obscuring the boundary 
between deterministic and nondeterministic effects. There is a better way. The 
nondeterministic character of the effects, with its attendant irreversibility, is 
the source of the problem, but it also provides the solution. In the same way that 
it prevents acquisition of precise information about the state of individual elementary 
systems, it also precludes detection of the \textit{sequence} of spacelike-separated 
nonlocal effects.

To see this consider two particles in an entangled state, $ \alpha|x_1\rangle|x_2\rangle +  \beta|y_1\rangle|y_2\rangle$, with $ |x\rangle$ and $|y\rangle$ orthogonal. Suppose that 
they separate to a macroscopic distance where they each encounter a series of 'detector' 
particles arranged to interact with the  $ |x\rangle$  branch. In \cite{Gillis_E} it was 
proposed that each such entangling interaction is accompanied by a small, probabilistic 
nonlocal shift of squared amplitude either from the interacting ($ |x\rangle$) branch to the 
noninteracting ($ |y\rangle$) branch of the wave function, or vice versa. Assume that the 
magnitude of the shifts of squared amplitude is 0.01. The state must be well defined 
at each stage of the process, so it is necessary to assume that the spacelike-separated 
interactions involving the left and right moving particles are sequenced in some fashion. 
Suppose that initially, we have $\alpha\alpha^* = \beta\beta^* = 0.5$. Imagine that after 
several hundred interactions have occurred on the left side, we take a "God's eye" view 
of the quantum state. If the sequencing of right and left side interactions is random, then 
we would expect that the nonlocal transfers from several hundred right side interactions 
would have also affected the state up to this point. Suppose that 
$\alpha\alpha^*$ is now $0.75$. On average 
this would mean that about $25^2 = 625$ transfers have occurred. The point is that this 
state is the \textit{only physical record of the sequence of nonlocal effects}, and it is 
perfectly consistent with over $2^{300}$ different sequences. If we were to add 
some other kinds of interactions to "watch" the process we would introduce additional 
entangling interactions with additional nondeterminstic effects. There is no way to 
record which one of the enormous number of possible sequences actually occurred. They 
are observationally equivalent. 

Relativity should be understood as a statement of this observational equivalence, and, 
to a large extent, as a consequence of the inherently nondeterministic nature of elementary 
interactions. The nondeterminism entails limits on the kinds of information that can be 
instantiated in physical systems, and so it regulates the ways in which information can be 
transmitted.  Relativity does \textit{not} rule out either nonlocal effects or the 
sequencing of those effects. It is an expression of the fact that there can be no 
physical record of such sequencing. 

Our view of the status of spacetime structure should be just as tentative and provisional 
as our attitude toward quantum wave functions. Introductory relativity texts typically 
talk about how reference frames are defined with respect to measuring rods and clocks.
In an address to the Prussian Academy of Sciences on January 27, 1921, Einstein said:
\begin{quotation}\noindent "All practical geometry is based upon...experience...
Suppose two marks have been put upon a practically-rigid rod. A pair of two such marks we 
shall call a tract. We imagine two practically-rigid bodies, each with a tract marked on 
it. These two tracts are said to be 'equal to one another' if the marks of the one tract 
can be brought to coincide permanently with the marks of the other."\end{quotation} 
He made a similar point about measurements of time with reference to clocks.

Compare this to a statement of Landau\cite{Landau} on interpreting the predictions of 
quantum theory:
\begin{quotation}\noindent "The ... quantitative description of the motion of an electron
requires the presence also of physical objects which obey classical mechanics to a sufficient 
degree of accuracy."\end{quotation} 

Now, restate Einstein's point in Landau's language:
\begin{quotation}\noindent "The ... [\textit{specification of a reference frame}] requires 
the presence also of physical objects which obey classical mechanics to a sufficient 
degree of accuracy."\end{quotation} 

The definition of reference frames and the interpretation of wave functions are dependent 
on the same physical objects, subject to the same physical laws. It is physical theory,  
taken as a whole, that is "relativistic" - not one particular aspect of that theory. 
Whether through genetic endowment or constant habituation, nature has equipped us with 
deep intuitions about how the world works. These intuitions, together with the historical 
path of discovery, have induced us to grant spacetime geometry a special status. Our 
intellectual apparatus is aimed, largely, at figuring out what we can control, so it is 
somewhat understandable that nature has not inclined us to more readily consider possible 
nondeterministic aspects of our world. But, in order to make current theory logically coherent,  
we need to realize that relativity is rooted as much in the indeterminism that characterizes 
quantum theory as in the structure of space and time.

\section*{Acknowledgements} 

Thanks to Brian E. Howard  for encouraging me to write this, and for reviewing it.

\end{document}